\documentclass[conference]{IEEEtran}
\IEEEoverridecommandlockouts
\usepackage{booktabs} 
\usepackage{array}
\newcolumntype{I}{!{\vrule width 1pt}}
\usepackage{paralist}
\usepackage{balance}
\usepackage{amssymb}
\usepackage{multirow}
\newcommand{\tabincell}[2]{\begin{tabular}{@{}#1@{}}#2\end{tabular}}
\usepackage{graphicx}
\usepackage{amsmath}
\usepackage{algorithm}
\usepackage{algorithmicx}
\usepackage{algpseudocode}

\algdef{SE}[DOWHILE]{Do}{doWhile}{\algorithmicdo}[1]{\algorithmicwhile\ #1}
\usepackage{cases}
\usepackage[english]{babel}  
\usepackage{color}

\begin{document}

\title{Learning-based Automatic Parameter Tuning for Big Data Analytics Frameworks}

\author{
\IEEEauthorblockN{Liang Bao$^{1}$, Xin Liu$^{2}$, Weizhao Chen$^{1}$}
\IEEEauthorblockA{$^{1}$School of Computer Science and Technology, XiDian University, Xi'an, Shaan Xi, China, 710071\\
$^{2}$Department of Computer Science, University of California, Davis, Davis, California, USA, 95616-8562\\
baoliang@mail.xidian.edu.cn, xinliu@ucdavis.edu, iknow123cwz@gmail.com}
}

\maketitle

\begin{abstract}
Big data analytics frameworks (BDAFs) have been \mbox{widely} used for data processing applications. These frameworks provide a large number of configuration parameters to users, which leads to a tuning issue that overwhelms users. To address this issue, many automatic tuning approaches have been proposed. However, it remains a critical challenge to generate enough samples in a high-dimensional parameter space within a time constraint. In this paper, we present AutoTune--an automatic parameter tuning system that aims to optimize application execution time on BDAFs. AutoTune first constructs a smaller-scale testbed from the production system so that it can generate more samples, and thus train a better prediction model, under a given time constraint. Furthermore, the AutoTune algorithm produces a set of samples that can provide a wide coverage over the high-dimensional parameter space, and searches for more promising configurations using the trained prediction model. AutoTune is implemented and evaluated using the Spark framework and HiBench benchmark deployed on a public cloud. Extensive experimental results illustrate that AutoTune improves on default configurations by 63.70\% on average, and on the five state-of-the-art tuning algorithms by 6\%-23\%.
\end{abstract}


\section{Introduction}\label{sec:intro}
Big data analytics frameworks (BDAFs), such as Hadoop MapReduce \cite{dean2008mapreduce}, Spark \cite{zaharia2012resilient}, and Dryad \cite{isard2007dryad}, have been increasingly utilized for a wide range of data processing applications. These applications have vastly diverse characteristics. To support such diversity, BDAFs provide a large number of parameters for users to configure. For example, both Hadoop and Spark have 100+ parameters that an application can configure \cite{dean2008mapreduce, spark2017configuration}. The configuration of these parameters significantly affects the application performance on BDAFs.

However, configuring such a large number of parameters is overwhelming to users \cite{zhu2017bestconfig}. As a result, users often accept the default settings \cite{ren2013hadoop}. Alternatively, manually tuning is widely adopted, which requires in-depth knowledge on both BDAFs and applications. It is labor-intensive, time-consuming, and often suboptimal. Therefore, there is a strong need for automatically tuning application configurations on BDAFs.

Because configuration parameters have high dimensionality, naive exhaustive search is not feasible. Existing \mbox{methods} include model-based, simulation-based, search-based, and \newline learning-based, as discussed in Section \ref{sec:related}. Among them, \newline learning-based tuning has received much recent attention. In general, it constructs a performance prediction model using training samples of different configurations, and then explores better configurations using some searching algorithms. Although previous studies on learning-based tuning show promising results, one critical challenge is to generate enough samples in a high-dimensional parameter space, because it is very time consuming. In practice, we often have time constraints on how long we can tune configurations.

To address this challenge, we present AutoTune--an automatic configuration tuning system that aims to tune \newline application-specific BDAF configurations within a given time constraint. AutoTune consists of two key components: AutoTune testbed and AutoTune algorithm. First, we construct a smaller-scale testbed on which we run most of the experiments under different configurations. The motivation is to obtain more samples in the high-dimensional parameter space so that we can train a better prediction model. The challenge is to construct a testbed that runs faster but still captures the performance variations of different configurations on the production system. Furthermore, the AutoTune algorithm searches for better configurations using both the testbed and the production system under the time constraint. The key is to generate a set of samples that can provide a wide coverage over the high-dimensional parameter space, and to search for more promising configurations using the trained model. It has to balance the effort on initialization, exploration and exploitation, and the best configuration selection on the production system.

In summary, our work makes the following contributions:
\begin{compactitem}

\item We propose a novel approach that derives a testbed to facilitate the exploration on the production system. It allows us to generate more training samples, and thus to produce a better prediction model, under a given time constraint.

\item We develop the AutoTune algorithm. It uses latin hypercube sampling (LHS) to generate effective samples in the high-dimensional parameter space, and multiple bound-and-search to select promising configurations in the bounded space suggested by the existing best configurations.

\item We evaluate the performance of AutoTune through extensive experiments using a well-known big data benchmark in a public cloud. We show that AutoTune \mbox{outperforms} default configurations by 63.70\% on average, and the five state-of-the-art tuning algorithms by 6\%-23\%.
\end{compactitem}

\section{Related Work} \label{sec:related}
Parameter tuning for BDAFs has received much attention from both industry and academia. Such work can be classified into four categories: model-based, simulation-based, search-based, and learning-based tuning.

In model-based tuning, analytical models are constructed based on domain knowledge a priori to predict and optimize the performance of BDAFs. Such approaches include the Starfish project \cite{herodotou2011starfish},  MR-COF \cite{liu2015mr}, MRTuner \cite{shi2014mrtuner}, etc. Model-based tuning relies on analysis for performance optimization, and thus can be done a priori without experiments. However, analytical models may fail to capture the highly complex runtime characteristics, especially as BDAFs evolve rapidly with new architectures and technologies.

Simulation-based tuning constructs combined simulation models that capture both the internal behavioral metrics of the BDAFs and the externally observed input-output relationships, such as in  \cite{kadirvel2012grey,wang2015performance,cardosa2011steamengine,wang2009simulation}. Such approaches need to determine all the factors that could affect performance, and to probe system internals many times to collect raw-data needed in the performance model.

Search-based tuning perceives parameter tuning problem as a black-box optimization problem and leverages a variety of search algorithms to explore good solutions, such as in BestConfig \cite{zhu2017bestconfig}, Gunther \cite{liao2013gunther}, MRONLINE \cite{li2014mronline}, and SPSA \cite{kumar2016performance}. Search-based tuning is easier to run and can be applied to general scenarios because it does not require any system-specific knowledge. However, it requires extensive experimentation on production systems, and thus time-consuming and sometimes impractical.

Most relevant to our work is learning-based tuning. Typically, one first constructs performance prediction models using training samples under different configurations, and then applies some searching algorithms to find better configurations based on these models. For instance, RFHOC uses random forests for performance prediction and a genetic algorithm to search for the Hadoop configuration space \cite{bei2016rfhoc}; ALOJA-ML \cite{berral2015aloja} identifies key performance properties of the workloads through several machine learning techniques, and predicts performance properties for a unseen workload; \cite{lama2012aroma} employed support vector machines (SVM) to predict the performance of Hadoop applications; \cite{yigitbasi2013towards} proposed a support vector regression (SVR) model; \cite{rizvandi2012modeling1} presented polynomial multivariate linear regression for MapReduce; \cite{tang2017system} used random forests and genetic algorithm; \cite{zhang2015finding} used a modified k-nearest neighbor algorithm to find desirable configurations based on similar past jobs that have performed well; \cite{kadirvel2012grey} compared twenty machine learning algorithms and identified four with high accuracy; \cite{luo2016performance} built SVM-based performance models for Spark using randomly modified and combined configurations; \cite{peng2017reinforcement} proposed a reinforcement learning approach; \cite{wu2013self} combined k-means++ clustering and simulated annealing algorithms; \cite{chen2015machine} employed an ensemble method with the combination of random sampling and hill climbing (RHC) method; and \cite{wang2016novel} considered multi-classification models.

In these approaches, it requires a large number of samples to construct a useful model. Generating such samples requires a significant amount of time on production systems \cite{bei2016rfhoc}, which is expensive or impractical. Thus we are motivated to address this key limitation.

Last, there are best practices based on industrial experience. For example, the official site on Spark tuning \cite{spark2017configuration} indicates that the data serialization and memory tuning are two main factors in tuning a Spark application. Other attempts include instrumentation \cite{wendell2013understanding}, tuning recommendations \cite{bida2016tuning}, and statistical analysis \cite{armbrust2013catalyst}.

\section{Problem Statement}
\begin{figure}
\centering
\includegraphics[width=0.35\textwidth]{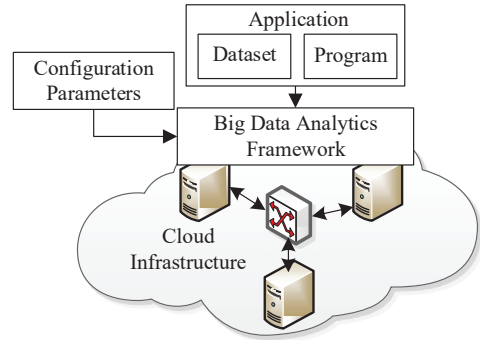}
\caption{Overview of PT-BDAF} \label{overview}
\end{figure}

In this work, we study the parameter tuning problem for big data analytics frameworks (PT-BDAF). As illustrated in Figure \ref{overview}, a big data analytics framework (e.g. Hadoop, Spark, etc.) is deployed on a collection of interconnected virtual machines (VMs) provided by a cloud provider. It serves data analysis applications comprised of programs and input datasets. In this process, the user who submits the application also needs to specify the configurations for the BDAF. Such configurations have a significant impact on application performance \cite{herodotou2011profiling, liu2015mr, zhu2017bestconfig, wang2015performance, bei2016rfhoc}.

The goal of PT-BDAF is thus to find an optimal configuration that minimize execution time, given a specific BDAF, an application, and the underlying runtime environment. Specially, PT-BDAF has the following components.

\textbf{Application}: An application represents a big data analytics task running on a specific BDAF. We model it as a 2-tuple $A$=$\left \langle p,d\right \rangle$, where $p$ is the program that expresses a set of computations; $d$ represents the input data to be processed by program $p$.

\textbf{Runtime Environment}: Runtime environment is the execution environment provided to a BDAF by the cloud infrastructure. We model it as a 5-tuple $E$=$\left \langle o,f,m,s,w\right \rangle$, where $o$ denotes the number of CPU cores; $f$ is the CPU frequency; $m$ represents the physical memory size; $s$ indicates the available disk space; and $w$ reflects the networking setup.

\textbf{Configuration and Execution Time}: Let $C$=$(c_{1},c_{2},\cdots$ $,c_{n})$ be the configuration of a BDAF. For example, in a Spark framework with 180+ parameters, an executor process can configure the number of cores and the memory size it uses, the maximum degree of parallelism, etc., as shown in Table \ref{tab:parameters}. Given a configuration $C$ for an application $A$ and its environment $E$, the execution time is denoted as $ET(A,E,C)$. In this paper, we treat $ET(\cdot)$ as a blackbox function to be learned.

\textbf{Time Constraint}: In practice, the time for configuration tuning is often restricted. We define this restricted tuning time as \emph{time constraint}, denoted as $TC$. Any solution to the \emph{PT-BDAF} problem must terminate when $TC$ is met.

In short, the \emph{PT-BDAF} problem can be stated as follows:
\begin{equation}
\min_{C \in \mbox{\footnotesize \emph{CB}}}ET(A,E,C)
\end{equation}
\begin{equation}
\mbox{\emph{s.t.}    tuning time} \leq TC
\end{equation}
where (1) states that the goal of \emph{PT-BDAF} problem is to find a configuration $C$ that minimizes execution time for a given application $A$ and its environment $E$. In $C$, the value of each component parameter $c_{i}$ must be within \emph{CB}, the configuration bound predefined by the BDAF. The constraint (2) is that any solution to the problem must terminate after a $TC$ amount of tuning time.

This definition shows that the goal of PT-BDAF is to search for an optimal configuration of a set of parameters to minimize the execution time. According to a previous study \cite{blum2003metaheuristics}, our PT-BDAF is essentially an instance of classic combinatorial optimization (CO) problems \cite{papadimitriou1982combinatorial}, which is known to be NP-complete. The NP-completeness proof by restriction is established in \cite{garey1979computers}.

Based on the above complexity analysis, we conclude that PT-BDAF is
NP-complete and non-approximable, which rules out the existence of
any polynomial-time optimal or approximate solution unless $P=NP$. Any complete methods that guarantees to find an optimal solution might need exponential computation time in the worst-case. This often leads to computation times too high for practical purposes \cite{blum2003metaheuristics}. Therefore, we shall focus on the design of a heuristic approach to this optimization problem.

\section{AutoTune for PT-BDAF}
\begin{figure}
\centering
\includegraphics[width=0.45\textwidth]{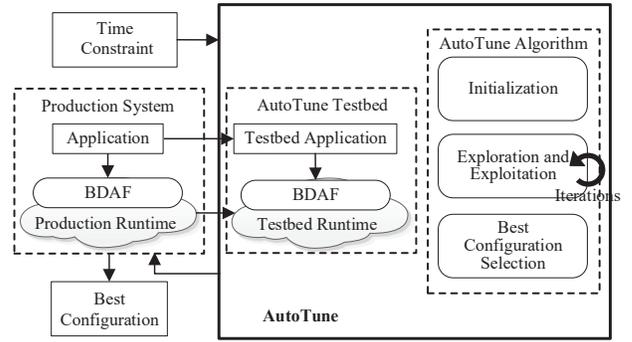}
\caption{Overview of AutoTune System} \label{fig:approach}
\end{figure}

In this section, we present AutoTune, a learning-based automatic parameter tuning system for BDAFs. Figure \ref{fig:approach} sketches the automatic parameter tuning process of AutoTune.

AutoTune consists of two important components: AutoTune testbed and AutoTune algorithm. The motivation of constructing a smaller-scale testbed instead of tuning directly on the production system is to obtain more samples to learn a better prediction model, and to conduct more iterations of exploration and exploitation to find a better configuration. The challenge is to construct a testbed that runs faster but still captures the performance variations of different configurations on the production system.

The AutoTune algorithm searches for better configurations by integrating the testbed and production system under the time constraint. The key of the AutoTune algorithm is to generate a set of samples that can provide a wide coverage over the high-dimensional parameter space, and search for more promising configurations using the trained model. It needs to balance the effort on the initialization, the exploration and exploitation, and the best configuration selection on the production system.

\subsection{AutoTune Testbed}
As discussed earlier, the most critical challenge in learning-based tuning is that obtaining training samples is time-consuming. To address this challenge, we propose a testbed-based approach. The goal of constructing a testbed is to evaluate the performance of different configurations on a BDAF in an accurate enough way but at a faster speed. The key is to reduce the size of dataset processed by the application and adjust the resource allocation properly so that the relative performance of different configurations on the testbed is as close as possible to that on the production system.

At a high level we consider a scenario where a user provides as input an analytics application (written using any existing data analytics framework) and a pointer to the input data for the application. Assuming that the machine types are fixed, we need to build a model first that will predict the execution time for any input size, number of machines for this given application. With the predictive model, the user can choose a appropriate testbed with certain reduction factor of the production system. Note that for the generality, we do not assume the presence of any historical logs about the application in order to infer the model.

The main steps in building such a predictive model are (a) capturing the computation properties of an application; (b) expressing the internal commination patterns in an application; and (c) determining how much data points we need to collect. We discuss all three aspects below.

\subsubsection{Computation properties}
As described in \cite{venkataraman2016ernest}, data analytics applications differ from other applications like SQL queries or stream processing in a number of ways. These applications are typically numerically intensive and thus are sensitive to the number of cores and memory bandwidth available. Further, these applications can also be long-running: for example, to obtain the state-of-the-art accuracy on tasks like image recognition and speech recognition, jobs are run for many hours or days.

\begin{figure}
\centering
\includegraphics[width=0.45\textwidth]{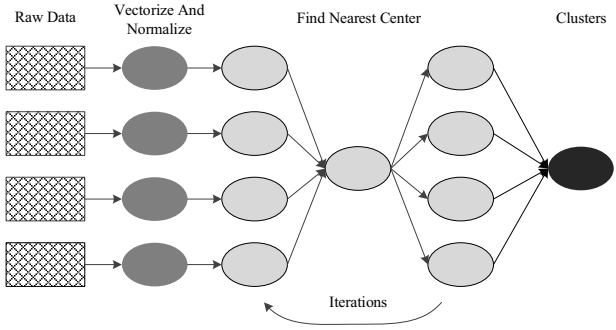}
\caption{Execution DAG of a well-known K-Means application.} \label{fig:kmeans}
\end{figure}

Take a well-known K-Means application, proposed in \cite{hartigan1979algorithm}, as an example of an data analytics application. The application divides points into clusters so that the within-cluster sum of squares is minimized, and its execution DAG is shown in Figure \ref{fig:kmeans}. From the figure we can see that this K-Means application contains three main stages: The first stage of the DAG reads input data, transforms the data into a collection of vectors, and normalizes each vector. The second stage in the application finds the nearest cluster centers by calculating the distance of each pair of vectors, and marks every vector with the nearest center. In the last stage, the vectors are grouped and represented by the centers, and the map-structured clustering results are returned to the user. The cluster centers are refined in every iteration and these steps are repeated for many iterations to achieve acceptable accuracy.

As observed in Figure \ref{fig:kmeans}, assuming the equal-dense input data and the equal-sized data partitions, we can find that each task in the application will take a similar amount of time to compute. Specifically, the computation required per data partition remains the same as we scale the input data, and if we add more machines to the cluster, the computation time decreases in linear or quasi-linear manner \cite{venkataraman2016ernest}.

\subsubsection{Communication patterns}
\begin{figure}
\centering
\includegraphics[width=0.45\textwidth]{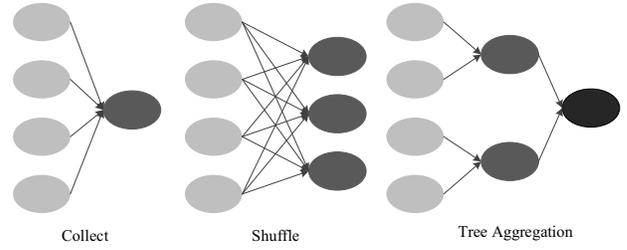}
\caption{Common communication patterns of BDAFs.} \label{fig:cpatterns}
\end{figure}

By investigating many state-of-the-art BDAFs, we observe that only a few communication patterns repeatedly occur in data analytics applications. These patterns (Figure \ref{fig:cpatterns}) include (a) the \emph{collect} (all-to-one) pattern, where data from all the partitions is sent to one machine; (c) the \emph{shuffle} (many-to-many) pattern where data goes from many source machines to many destinations; and (c) the \emph{tree-aggregation} pattern where data is aggregated using a tree-like structure. Actually, these patterns are not specific to analytics applications and have been wildly studied in many different distributed computing frameworks \cite{gropp1999using, chowdhury2012coflow}. Having a handful of such patterns means that we can try to automatically infer how the communication costs change as we increase the scale of computation. For example, assuming that data grows as we add more machines (i.e., the data per machine is constant), the time taken for the collect increases as $O$(\emph{nm}) as a single machine needs to receive all the data, similarly the time taken for a binary aggregation tree grows as $O(log$(\emph{nm})), where $nm$ represents the number of machines.

\subsubsection{Predictive model}
To build our model, we add terms related to the computation and communication patterns as discussed earlier. Specifically, based on \cite{venkataraman2016ernest}, we propose the following model:
\begin{equation}\label{testbedpredictionmodel}
t=[\theta_{0} + \theta_{1} \times \frac{ds}{nm}]+[\theta_{2}\cdot log(nm)+\theta_{3}\cdot nm]
\end{equation}
where $ds$ represents the data scale of the input data size, $nm$ denotes the number of machines. As shown in Eq.(3), the terms we add to our linear model are:
\begin{compactitem}
\item The first term of Eq.(\ref{testbedpredictionmodel}) consists of a fixed cost term that represents the amount of time spent in serial computation, and the interaction between the data size and the inverse of the number of machines. This term is to capture the parallel computation time for tasks, i.e., if we double the number of machines with the same size of input data, the the computation time will reduce linearly.

\item The second term contains a $log(nm)$ term to model communication patterns like tree-aggregation trees, and a linear term $O(nm)$ which captures the all-to-one communication pattern and fixed overheads like scheduling/serializing tasks (i.e. overheads that scale as we add more machines to the system).
\end{compactitem}

Note that as we use a linear combination of non-linear features in Eq.(\ref{testbedpredictionmodel}), we can model non-linear behavior as well.

The objective of the training is to learn values of $\theta_{0}$, $\theta_{1}$, $\theta_{2}$, and $\theta_{3}$. We can use a non-negative least square (NNLS) \cite{lawson1995solving} solver to find the model that best fits the training data. NNLS fits our use case very well as it ensures that each term contributes some non-negative amount to the overall time taken. This avoids over-fitting and also avoids corner cases where say the running time could become negative as we increase the number of machines.

\subsubsection{Projective sampling}
The next step is to collect training samples for building our predictive model. Specifically, we use the input data provided by the user and run the complete job on small samples of the data and collect the time taken for the application to execute. The ultimate goal of this step is minimizing the time spent on collecting training data while achieving good enough accuracy.

To improve the time taken for training without sacrificing the prediction accuracy, we outline a scheme based on \emph{projective sampling} \cite{last2009improving}, a state-of-the-art technique that fits a function to a partial learning curve obtained from a small subset of potentially available data and then uses it to analytically estimate the optimal training set size. More specifically, based on the common assumption that the error rate is a non-increasing function of the sample size $n$ \cite{provost1999efficient}, we use projective sampling to predict the number of samples required to build the predictive model. The initial samples are selected by randomly adding a constant number of samples (combinations of $ds$ and $nm$) to the training set from the training pool. In each iteration, the model is built, the accuracy of the model is evaluated using the testing data, and a sample point for the learning curve is thus generated. Using the information from the already generated sample points, we follow the approach proposed by Last \cite{last2009improving} in selecting the known projective learning function (including Logarithmic, Weiss and Tian, Power Law and Exponential functions) that exhibits highest correlation with these points. Once we have determined the best-fit function, we can calculate the optimal sample size that ensures the most optimal tradeoff between sampling cost and prediction accuracy.

\subsubsection{Testbed construction}
With the predictive model and the sampling strategy, we can now discuss the testbed construction algorithm in Algorithm \ref{autotunetestbed}.

\begin{algorithm}
\small
\caption{\textbf{\emph{AutoTuneTestbed(PS, TC, f, RC)}}}
\begin{algorithmic}[1]
\Require \emph{PS}: the production system; \emph{TC}: time constraint; \emph{f}: scale factor; \emph{RC}: resource constraint.

\State $n \leftarrow 0$; //Initialize the training set size $n$

\State $i \leftarrow 0$; //Initialize the number of data points used to compute
the equation of the learning curve

\Do

\State Acquire $\delta$ samples with different $ds$ and $nm$ values;

\State $n \leftarrow n + \delta$;

\State $i \leftarrow i + 1$;

\State Run applications under $\delta$ different settings, collect execution time results, and update the training set \emph{T};

\State Train the $i$-th prediction model $M_{i}$ using Eq.(3) with \emph{T};

\State $Best\_Corr \leftarrow 0$;

\State Calculate the correlation coefficient $Corr_{j}^{i}$ for each learning curve function $j$ based on the $M_{i}$;

\State $Best\_Corr \leftarrow min(Corr^{i}_{j})$;

\State $Best\_Function \leftarrow$ the best functional form found so far;

\doWhile{(\emph{TC} permits more tests) \textbf{and} ($Best\_Corr \geq 0$)}

\State Estimate the optimal training set size $n^{*}$ according to the selected $Best\_Function$;

\While {(\emph{TC} permits more tests) \textbf{and} ($n < n^{*} $)}

\State $n \leftarrow n + \delta$;

\State $i \leftarrow i + 1$;

\State Run application under $\delta$ different settings, collect execution time results, and update the training set \emph{T};

\State Train the $i$-th prediction model $M_{i}$ using Eq.(3) with \emph{T};

\EndWhile

\State Run application on the \emph{PS} with default configuration $C_{0}$ and estimate the execution time $t_{0}$;

\State \Return All possible testbed settings of $ds$ and $nm$, each of which has the predictive execution time $t=t_{0}\cdot f$, and must satisfy the resource constraint \emph{RC};

\end{algorithmic}
\label{autotunetestbed}
\end{algorithm}

The Autotune testbed algorithm starts with an empty training set, adding a constant number $\delta$ of samples to the training set in each iteration (line 4). Once the samples are selected, the corresponding performance values are evaluated (line 7). The samples and the associated performance values are then used to build a predictive model for the testbed (line 8).

Each iteration adds a sample point for the calculation of the learning curve equation, where the cumulative training set size $n$ is considered an independent variable and the error rate of the model induced from $n$ examples is treated as the dependent variable. Given the newly learned model $M_{i}$, we use Pearson's correlation coefficient \cite{pearson1895note} to estimate the correlation for each candidate learning curve function (lines 10--12). Since in a ``well-behaved'' learning curve, the error rate should be a non-decreasing function of $n$, a function with a minimal (closest to $-1$) negative correlation coefficient should be the best fit for the data. However, the actual data points may be noisy, occasionally resulting in positive correlation coefficients over a limited number of data points. In that case, the algorithm will keep purchasing additional samples until the lowest correlation coefficient becomes negative or the maximum amount of available training examples is exceeded (line 13).

Once we have determined the best-fit function that can approximate the learning curve accurately, we can calculate the coefficients of the projected function using the least-squares method \cite{marquardt1963algorithm} and estimate the optimal training set size $n^{*}$ (line 14). If $n^{*}$ is greater than the size of the current training set and the \emph{TC} permits more tests, the algorithm purchases the missing amount of examples; otherwise the algorithm stops and update the predictive model (lines 15--20).

With the final predictive model, the algorithm returns all possible testbed settings, each of which has the expected execution time of scale factor $f$, and satisfies the resource constraint \emph{RC} proposed by user (lines 21--22).

Note that the fixed sampling increment $\delta$ in Algorithm \ref{autotunetestbed} is determined in advance based on domain-specific constraints such as the minimum number of batches in each experiment. We set the value of $\delta$ to 5 in our experiments, and have seen from experiments that even a value of 1 can give good results \cite{sarkar2015cost}.

\subsection{AutoTune Algorithm}
We now discuss the AutoTune algorithm in Algorithm \ref{algorithm}.

\begin{figure}
\centering
\includegraphics[width=0.2\textwidth]{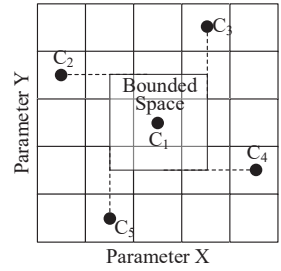}
\caption{An example of bounding mechanism for a 2D space} \label{fig:bound-and-search}
\end{figure}

\begin{algorithm}
\small
\caption{\textbf{\emph{AutoTuneAlgorithm(TB,C,CB,TC)}}}
\begin{algorithmic}[1]
\Require \emph{TB}: the testbed; \emph{C}: configuration; \emph{CB}: configuration bounds; \emph{TC}: time constraint.

\State Generate $h$ different configs using LHS within \emph{CB};

\State Run these $h$ configs on \emph{TB}, collect execution time results, and generate the initial training set \emph{T};

\State $B \leftarrow$ the best $b$ configs in \emph{T};

\While {\emph{TC} permits more tests}

\State Generate $h$ different configs using LHS within \emph{CB}, and then choose $b$ out of $h$ configs randomly;

\State Run these $b$ configs on \emph{TB}, collect execution time results, and generate the exploration set \emph{EP};

\State $T\leftarrow T \bigcup$\,\emph{EP};

\State Train a prediction model $M$ using \emph{T};

\For {each config $C_{i} \in B$}

\State Set exploitation set \emph{EI}$\leftarrow\varnothing$;

\State Select $h$ configs in the bounded space around $C_{i}$;

\State Predict the execution time results of these $h$ configs using $M$, and choose the best config $C_{i}^{*}$;

\State Run \emph{TB} with $C_{i}^{*}$, collect the execution time $t_{i}^{*}$;

\State \emph{EI}=\emph{EI}\,$\bigcup \{(C_{i}^{*},t_{i}^{*})\}$;

\EndFor

\State $B\leftarrow$ the best $b$ configs from $B\bigcup$\,\emph{EP}\,$\bigcup$\,\emph{EI};

\State $T\leftarrow T \bigcup$\,\emph{EI};

\EndWhile

\State Run the best $q$ configs in $B$ on the production system and select the best one;

\State \Return The best config;

\end{algorithmic}
\label{algorithm}
\end{algorithm}
\textbf{1. Latin Hypercube Sampling}. A key component in AutoTune algorithm is its sampling strategy. Because the configuration parameter space is high dimensional, it is challenging to provide a good coverage in it, especially with a low number of samples. In such a scenario, latin hypercube sampling (LHS) performs better, compared to random or grid sampling, because it allows each of the key parameters to be represented in a fully stratified manner, no matter which parameters are important \cite{mckay1979comparison}. Specifically, LHS divides the range of each parameter into $h$ intervals and take only one sample from each interval with equal probabilities \cite{mckay1979comparison}. The general LHS algorithm for generating $h$ random vectors (or configurations) of dimension $n$ can be summarized as follows:
\begin{compactitem}
\item [1)] Generate $n$ random permutations with $h$-dimensions $\{1,2,\dots,h\}$, denoted by $\vec{P^{1}}$, $\vec{P^{2}}$,\dots, $\vec{P^{n}}$ where $\vec{P^{i}}=(P_{1}^{i},P_{2}^{i},\cdots,P_{h}^{i})$.
\item [2)] For the $i$-th dimension ($i$=$1,2,\cdots,n$), divide the parameter range $c_{i}$ into $h$ non-overlapping intervals of equal probabilities.
\item [3)] The $j$-th sampled point is an $n$ dimensional vector, with the
value for dimension $i$ uniformly drawn from the $P_{j}^{i}$-th interval of $c_{i}$.
\end{compactitem}
Figure \ref{fig:bound-and-search} illustrates an example of LHS with five intervals in a 2D dimension, where $C_{1}$-$C_{5}$ denote the five LHS samples. Note that a set of LHS sample with $h$ vectors will have exactly one point in every interval on each dimension. That is, LHS attempts to provide a coverage of the experimental space as evenly as possible. Compared to pure random sampling, LHS provides a better coverage of the parameter space and allows a significant reduction in the sample size to achieve a given level of confidence without compromising the overall quality of the analysis \cite{xi2004smart}.

AutoTune uses LHS for sampling (line 1 and 5). Line 1-3 generates the initial training set, where $h$ is a hyperparameter, discussed later.

\textbf{2. Training a Prediction Model}. Based on these samples, we have tried different machine learning algorithms to train a prediction model. Specifically, random forests achieve good performance and thus is adopted (line 8).

\textbf{3. Exploration and Exploitation}. To explore the parameter space, we apply LHS again with $h$ intervals and choose $b$ configurations randomly (line 5). After that, we run these $b$ configurations on the testbed and use the results to generate the exploration set \emph{EP} (line 6).

To exploit the previously-found best configurations, we design a two-step multiple bound and search (MBS) algorithm to find potential better configurations near already-known good configurations (line 9-15). This strategy works well in practice because there is a high possibility that one can find other configurations with similar or better performances around the configuration with the best performance in the sample set \cite{zhu2017bestconfig}.

\textbf{Bound-and-sample}. For each configuration $C_{i}$ in $B$, MBS generates another set of samples in the bounded space around $C_{i}$. The bounded space is generated as follows. For each parameter $c_{j}$ in $C_{i}$, MBS finds the largest value $c_{j}^{l}$ (lower bound) that is represented in $B$ and is smaller than that of $C_{i}$. It also finds the smallest value $c_{j}^{u}$ (upper bound) that is represented in $C_{i}$ and that is larger than that of $C_{i}$. The same bounding mechanism are carried out for every $c_{j}, j=1,2,\cdots,n$ in $C_{i}$. Figure \ref{fig:bound-and-search} illustrates this bounding mechanism of MBS with 2D space. After determining the bound for $C_{i}$, we use LHS again to divide each bound into $h$ intervals and generate $h$ samples close to $C_{i}$ (line 11).

\textbf{Search}. Given these $h$ configurations, we use the trained prediction model $M$ to choose the best configuration $C_{i}^{*}$ (line 12). We then run the testbed with $C_{i}^{*}$ and collect the corresponding execution time $t_{i}^{*}$ (line 13). Last, the sample ($C_{i}^{*},t_{i}^{*}$) is added to the exploitation set $EI$ (line 14).

We repeat these two steps until every configuration in $B$ is tested, and update $B$ with the best $b$ configurations from $B\bigcup$\,\emph{EP}\,$\bigcup$\,\emph{EI} (line 16). We refine $M$ by adding new samples from exploration and exploitation phases to the training set $T$ (line 7 and 17).

\textbf{4. Selecting the best configuration}. Once the time budget on exploration and exploitation is met, we stop searching and run the best $q$ configurations in $B$ on the production system (line 19). Finally we return the best one from these $q$ configurations.

To satisfy the overall time constraint, we divide it into three phases, i.e. initialization, exploration and exploitation, and the best configuration selection. Suppose the time constraint is $TC$, the proportion of time spent in these three phases is denoted as $\alpha$, $\beta$, and $\gamma$, respectively, where $\alpha+\beta+\gamma=1$ and $0 \leq \alpha,\beta,\gamma \leq 1$. Let the average time of running an application with one configuration on testbed equals to $t_{TB}$, and the average time on production system is $t_{PS}$, we have $h\approx\lfloor \frac{\alpha*TC}{t_{TB}} \rfloor$, $q\approx\lfloor \frac{\beta*TC}{t_{PC}} \rfloor$, and $b\approx\lfloor \frac{\gamma*TC}{\mbox{\scriptsize\emph{iter}}*t_{TB}} \rfloor$, where \emph{iter} represents the approximate iterations in the exploration and exploitation phase, and $h$, $b$, $q$ are three hyperparameters in AutoTune algorithm. We will discuss the performance variations over different time ratios in Section \ref{sec:experimentresults}.

\section{Experiments}
\subsection{Experimental Settings}
\textit{Runtime environment}. We conduct our experiments on a public cloud infrastructure named Aliyun\footnote{https://www.aliyun.com}. We use 6 Aliyun ECS instances consisting of two types: 1 general type instance (g5) for a master node, and 5 memory type (r5) for slave nodes. The master node is equipped with a 4-core Intel Skylake Xeon Platinum 8163 2.5GHz processor, 16GB RAM, and 100G disk. Each of the slave nodes is equipped with a 4-core Intel Skylake Xeon Platinum 8163 2.5GHz processor, 32GB RAM, and 250G disk. Both instances have CentOS 6.8 (64bit) installed. All of the VM instances are connected via a high-speed 1.5Gbps LAN.

\textit{Framework and configuration setup}. We choose Spark as our experimental framework. Spark is a general-purpose cluster computing engine for streaming, graph processing and machine learning \cite{zaharia2012resilient}. We choose Spark because it is a widely adopted open-source data processing engines.

Based on the Spark manual \cite{spark2017configuration} and previous studies \cite{zhu2017bestconfig, luo2016performance, wang2016novel, wang2015performance}, we identify 13 out of 180+ parameters that are considered critical to the performance of Spark applications, as listed in Table \ref{tab:parameters}. It's worth noting that even with only 13 parameters, the search space is still enormous, and exhaustive search is infeasible.

\begin{table*} \small
\centering \caption{Performance-aware parameters in Spark}
\label{tab:parameters}
\begin{tabular}{|l|l|l|}\hline

\textbf{Parameters} & \textbf{Brief Description} &\tabincell{l}{\textbf{Default Value}}\\
\hline
spark.executor.cores &\tabincell{l}{The number of
cores to use on each executor} & 4 \\ \hline
spark.executor.memory & \tabincell{l}{Amount of memory to use per
executor process}& 1024MB\\ \hline
spark.memory.fraction & \tabincell{l}{Fraction of heap space used for execution and storage} & 0.6 \\ \hline
spark.memory.storageFraction & \tabincell{l}{Amount of storage memory immune to eviction, \\ expressed as a fraction of the size of \\ the region set aside by spark.memory.fraction} & 0.5 \\ \hline
spark.default.parallelism & \tabincell{l}{Default number of partitions in RDDs \\ returned by transformations} & 20 \\ \hline
spark.shuffle.compress & \tabincell{l}{Whether to compress map output files} & True \\ \hline
spark.shuffle.spill.compress & \tabincell{l}{Whether to compress data spilled during shuffles} & True \\ \hline
spark.broadcast.compress & \tabincell{l}{Whether to compress broadcast variables before sending them} & True \\ \hline
spark.rdd.compress & \tabincell{l}{Whether to compress serialized RDD partitions} & False \\ \hline
spark.io.compression.codec & \tabincell{l}{The codec used to compress internal data such as RDD partitions, \\ event log, broadcast variables and shuffle outputs} & lz4 \\ \hline
spark.reducer.maxSizeInFlight & \tabincell{l}{Maximum size of map outputs to fetch simultaneously \\ from each reduce task (MB)} & 48MB \\ \hline
spark.shuffle.file.buffer & \tabincell{l}{Size of the in-memory buffer for each shuffle file \\ output stream (KB)} & 32KB \\ \hline
spark.serializer & \tabincell{l}{Class to use for serializing objects that will be sent \\ over the network or need to be cached in serialized form} & JavaSerializer \\ \hline
\end{tabular}
\end{table*}

\textit{Benchmark}. The HiBench \cite{huang2010hibench} benchmark is used in our experiments. We select seven representative applications in three different categories: micro benchmark, machine learning, and websearch benchmark. Table \ref{tab:benchmark} lists the applications and the corresponding dataset size.

\begin{table}\small
\centering \caption{Applications and corresponding dataset size}
\label{tab:benchmark}
\begin{tabular}{|c|l|c|}\hline
\textbf{Abbr.} & \textbf{Program} & \textbf{Dataset Size} \\
\hline

WC & WordCount & 76.5GB \\ \hline

BC & Bayesian classification & 5.6GB \\ \hline

KC & K-Means clustering & 38.3GB \\ \hline

LR & Logistic regression & 7.5GB \\ \hline

SVM & Support vector machine & 80.8GB \\ \hline

GBT & Gradient boosting trees & 603.2MB \\ \hline

PR & PageRank & 506.9MB  \\ \hline

\end{tabular}
\end{table}

\subsection{Baseline Algorithms}
To evaluate the performance of AutoTune, we compare it with five state-of-the-art algorithms, namely random search \cite{bergstra2012random}, BestConfig \cite{zhu2017bestconfig}, RFHOC \cite{bei2016rfhoc}, Hyperopt \cite{bergstra2013hyperopt}, and SMAC \cite{hutter2011sequential}. We provide a brief description for each algorithm and report its hyperparameters (if necessary) as follows:

\textbf{Random search (Random)} is a search-based tuning approach that explores each dimension of parameters uniformly at random. It is more efficient than grid search in high-dimensional configuration spaces, and is a high-performance baseline, as suggested in \cite{bergstra2012random}.

\textbf{BestConfig}\footnote{Code is available from: https://github.com/zhuyuqing/bestconf} is a search-based tuning approach that uses divide-and-diverge sampling and recursive bound-and-search algorithm to find a best configuration. We follow the suggestions in \cite{zhu2017bestconfig} and tabulate the value of $k$ in Table \ref{tab:controlparameters}.

\textbf{RFHOC} is a learning-based tuning approach that constructs a prediction model using random forests, and a genetic algorithm to automatically explore the configuration space. We use the hyperparameters suggested in \cite{bei2016rfhoc}.

\textbf{Hyperopt}\footnote{Code is available from: http://jaberg.github.io/hyperopt/} is a learning-based tuning approach based on Bayesian optimization. It is widely used for hyperparameter optimization. We use the suggested settings in \cite{bergstra2013hyperopt}.

\textbf{SMAC}\footnote{Code is available from: http://www.cs.ubc.ca/labs/beta/Projects/SMAC} is a learning-based tuning method using random forests and an aggressive racing strategy.

In AutoTune algorithm, we set the number of iterations to $100$ for random forest model, and do not limit the depth of the decision tree and the number of available features for the tree; the values of three hyperparameters, i.e. $h$, $b$, and $q$, for each application are listed in Table \ref{tab:controlparameters}.

For each run in our experiments, every algorithm is executed under the same time constraint and stops once the constraint is met.

\begin{table}\small
\centering \caption{Hyperparameters in BestConfig and AutoTune}
\label{tab:controlparameters}
\begin{tabular}{|c|c|c|c|c|c|c|c|}\hline
\multirow{3}*{\textbf{Apps}} & \multicolumn{2}{c}{\textbf{BestConfig}} & \multicolumn{5}{|c|}{\textbf{AutoTune}} \\ \cline{2-8}

& PS & TB & \multicolumn{2}{c|}{PS} & \multicolumn{3}{c|}{TB} \\ \cline{2-8}

& $k$ & $k$ & $h$ & $b$ & $h$ & $b$ & $q$ \\ \hline

WC & 31 & 211 & 31 & 5 & 420 & 10 & 29 \\ \hline

BC & 100 & 446 & 100 & 5 & 446 & 10 & 61 \\ \hline

KC & 25 & 324 & 25 & 5 & 320 & 20 & 51\\ \hline

LR & 20 & 163 & 20 & 5 & 163 & 10 & 24\\ \hline

SVM & 9 & 88 & 9 & 2 & 88 & 10 & 15\\ \hline

GBT & 23 & 64 & 23 & 5 & 64 & 10 & 15 \\ \hline

PR & 79 & 317 & 79 & 5 & 317 & 10 & 50 \\ \hline

\end{tabular}
\end{table}

\subsection{Evaluation Metrics}
We consider three performance metrics in our experiments for performance evaluation, namely cost of testbed construction, nDCG and ET. Cost of testbed construction measures the cost of building a testbed, nDCG is a metric that evaluates the quality of a testbed, and execution time is the ultimate performance metric for AutoTune.

\textbf{Cost of testbed construction}. The critical step of AutoTune testbed is to derive a performance prediction model that can guide the construction of testbed under the time constraint and the desired scale factor proposed by users. Typically, performance prediction models are evaluated on the basis of their prediction accuracy. It is also common knowledge that usually a larger training set results in higher prediction accuracy \cite{sarkar2015cost}. However, a large training set is often undesirable in terms of measurement effort in this problem. Thus, any performance prediction model built for this purpose should be evaluated not only in terms of prediction accuracy, but also in terms of measurement cost involved in building the training and testing sets. More specifically, we adopt the cost model proposed in \cite{sarkar2015cost} to include the cost incurred in measuring the testing set along with the training set:
\begin{equation}\label{eq:costfunction}
  TotalCost(n)=2n + \epsilon_{n} \cdot|S| \cdot R
\end{equation}
where $2n$ is the number of samples in the training ($n$ samples) and testing sets ($n$ samples), $\epsilon_{n}$ is the prediction error of the predictive model for testbed built with the $n$ samples, $|S|$ represents the number of settings whose performance value will be predicted by the model, and $R=1$ means that we equally weigh the cost incurred in measuring samples and the cost due to prediction error \cite{sarkar2015cost}. Note that in this definition, we ignore the cost incurred in building a performance prediction model for the testbed, as for linear regression model (Eq.(\ref{testbedpredictionmodel})), which is used in our approach, this cost is computationally insignificant, compared to the other cost factors.

\textbf{Normalized Discounted Cumulative Gain (nDCG)} is originally used to evaluate a ranking-quality metric of a search result set \cite{jarvelin2002cumulated}. We use it here to evaluate the quality of a ranking for a set of configurations generated from the testbed by comparing it with the corresponding real ranking from the production system. Note that the ranking of a prediction model is more important than its prediction accuracy in AutoTune, because we select the subset of best configurations in AutoTune and feed it into the production system (line 19 in Algorithm \ref{algorithm}).

Specially, given a ranking $r$ with $n$ configurations, its DCG is defined as: $$DCG_{r}=\sum_{i=1}^{n}\frac{2^{rel_{i}}-1}{log_{2}(i+1)},$$ where $rel_{i}$ is the graded relevance of the result at position $i$. Given a predicted ranking $r_{i}$ and its true ranking $r_{i}^{*}$ at position $i$, we define a 5-level relevance rating criteria by calculating the absolute relative deviation between $r_{i}$ and $r_{i}^{*}$, as shown in Table \ref{tab:relevance}.

The normalized DCG of a ranking $r$ is thus defined as the ratio of $DCG_{r}$ to the real ranking sequence $DCG_{r^{*}}$: $$nDCG_{r}=\frac{DCG_{r}}{DCG_{r^{*}}}$$

For example, suppose the real ranking $r^{*}$ of three configurations is $(1, 2, 3)$, and the predicted ranking $r$ of these settings is $(2, 1, 3)$, the DCG of $r^{*}$ is $66.03=31+\frac{31}{log_{2}(3)}+\frac{31}{2}$; the DCG of $r$ is $26.91=7+\frac{7}{log_{2}(3)}+\frac{31}{2}$, since the first and the second relevance values are both 3 (good), and the last relevance value is 5 (perfect). The nDCG of $r$ is thus equal to $0.41=\frac{26.91}{66.03}$.

\begin{table}\small
\centering \caption{The definition of a 5-level relevance values}
\label{tab:relevance}
\begin{tabular}{|c|c|c|}\hline
\textbf{Relevance Rating} & \textbf{Value} & \textbf{Condition} \\
\hline

Perfect & 5 & $\frac{|r_{i}-r_{i}^{*}|}{n-1}\leq 0.1$ \\ \hline

Excellent & 4 & $0.1 < \frac{|r_{i}-r_{i}^{*}|}{n-1}\leq 0.25$ \\ \hline

Good & 3 & $0.25 < \frac{|r_{i}-r_{i}^{*}|}{n-1}\leq 0.55$ \\ \hline

Fair & 2 & $0.55 < \frac{|r_{i}-r_{i}^{*}|}{n-1}\leq 0.9$ \\ \hline

Bad & 0 & $0.9 < \frac{|r_{i}-r_{i}^{*}|}{n-1}\leq 1$ \\ \hline

\end{tabular}
\end{table}

\textbf{Execution Time (ET)} is the total completion time of the application from the time when the request is fed into the BDAF to the time when the result is returned to the client. The ET improvement of an algorithms over the baseline algorithm in comparison is defined
as:$$Imp(baseline)=\frac{ET-ET_{baseline}}{ET_{baseline}} \times 100\%,$$ where $ET_{baseline}$ is the execution time of the baseline, and $ET$ is that of the algorithm being evaluated.

To ensure consistency, we run each application five times and calculate the average of these five runs. The standard deviation of the execution time is 0.02 or smaller, which indicates the stability of the performance.

\subsection{Experiment Results}\label{sec:experimentresults}
\subsubsection{Testbed evaluation} The first experiment is to evaluate the effectiveness of our AutoTune testbed approach. By conducting experiments on seven Spark applications, we aim at answering the following two research questions:
\begin{itemize}
  \item RQ1: Is the projective sampling strategy in our AutoTune testbed algorithm more cost efficient than classic progressive sampling method?
  \item RQ2: How about the quality of testbeds with different scale factors?
\end{itemize}

Since cost efficiency is the primary determinant for judging the effectiveness of a sampling strategy, we compared the total cost of sampling and prediction according to Eq.(\ref{eq:costfunction}), for all seven applications, using progressive and projective sampling. Progressive sampling is a popular sampling strategy that has been used for a variety of learning models. The central idea is to use a sampling schedule $n_{0}, n_{1}, \dots, n_{k}$, where each $n_{i}$ is an integer that specifies the size of the sample set that is used to build a performance prediction model at iteration $i$. In our experiment, we adopt the geometric progressive sampling strategy, where $n_{i}=n_{0} * a^{i}$. The parameter $a$ is a constant that defines how fast we increase the size of the sample set.

As shown in Table \ref{tab:samplingcompare}, for both progressive and projective sampling, we calculated the cost and accuracy of building prediction models with the optimal sample set size ($n^{*}$). The value of $n^{*}$ is determined through the respective sampling techniques. We see in Table \ref{tab:samplingcompare} that projective sampling outperforms progressive sampling in terms of cost and also in terms of accuracy. For SVM and GBT, progressive sampling gets stuck in a local optimum and produces low accuracies of 20\% and 15\%. Progressive and projective sampling are comparable in terms of accuracy and cost for WC. However, for all other four applications, projective sampling is considerably more cost efficient than progressive sampling.

\begin{table} \small
\centering \caption{Cost and accuracy of progressive and projective sampling}
\label{tab:samplingcompare}
\begin{tabular}{|c|c|c|c|c|}\hline
\multirow{2}*{\textbf{Apps}} & \multicolumn{2}{c|}{\textbf{Cost}} &
\multicolumn{2}{c|}{\textbf{Accuracy (\%)}} \\ \cline{2-5}

& Progressive & Projective & Progressive & Projective \\ \hline

WC & 220 & \textbf{180} & 85 & \textbf{88} \\ \hline

BC & 877 & \textbf{380} & 90 & \textbf{92} \\ \hline

KC & 1235 & \textbf{423} & 89 & \textbf{89} \\ \hline

LR & 1492 & \textbf{516} & 87 & \textbf{85} \\ \hline

SVM & 276 & \textbf{89} & 20 & \textbf{91} \\ \hline

GBT & 78 & \textbf{42} & 15 & \textbf{83} \\ \hline

PR & 415 & \textbf{231} & 78 & \textbf{86} \\ \hline

\end{tabular}
\end{table}

To measure the quality of testbeds, we need to evaluate the nDCG values on testbeds with different scale factors. For each application, we construct the testbeds with many combinations of data scale ($ds$) and the number of machines ($nm$). After that, we randomly generate 30 different configurations, and collect the execution time on both production system ($PS$) and different testbeds ($TB$s). Finally, we calculate the nDCG values of these $TB$s. Table \ref{tab:testbedquality} lists the nDCG values of the K-Means application with ten $ds$ values and five $nm$ values.

\begin{table*} \small
\centering \caption{The nDCG values of K-Means application with different settings of testbeds}
\label{tab:testbedquality}
\begin{tabular}{|c|c|c|c|c|c|c|c|c|c|c|}\hline
\multirow{2}*{\tabincell{c}{\textbf{\# of} \\ \textbf{machines}}} & \multicolumn{10}{c|}{\textbf{nDCG values with different data scale}} \\ \cline{2-11}
& \textbf{0.03125} & \textbf{0.05} & \textbf{0.0625} & \textbf{0.1} & \textbf{0.125} & \textbf{0.25} & \textbf{0.5} & \textbf{0.6} & \textbf{0.8} & \textbf{0.9}\\ \hline

\textbf{1} & 0.4407 & 0.5288 & 0.6091 & 0.5929 & 0.5991 & 0.6032 & 0.6432 & 0.6054 & 0.5928 & 0.7002 \\ \hline

\textbf{2} & 0.4252 & 0.5239 & 0.7070 & 0.7158 & 0.6229 & 0.6774 & 0.7675 & 0.6743 & 0.7863 & 0.8162 \\ \hline

\textbf{3} & 0.4213 & 0.5032 & 0.7872 & 0.7865 & 0.7923 & 0.6258 & 0.6632 & 0.7058 & 0.7721 & 0.8765 \\ \hline

\textbf{4} & 0.4726 & 0.5911 & 0.8543 & 0.8170 & 0.7662 & 0.8345 & 0.8732 & 0.8251 & 0.9021 & 0.9854 \\ \hline

\textbf{5} & 0.4942 & 0.6032 & 0.9462 & 0.9352 & 0.8621 & 0.9954 & 0.9007 & 0.9976 & 0.9171 & 0.9986 \\ \hline

\end{tabular}

\end{table*}

We can see from Table \ref{tab:testbedquality} that the testbeds having the same number of machines ($nm$) as the production system obtain better nDCG values than others. This is because changing the number of machines in the runtime environment will lead to corresponding changes on resource provision and scheduling, it can cause unpredictable performance of the application. Therefore, we should keep the underlaying environment of testbed as close as possible to the production system. Another interesting result found in Table \ref{tab:testbedquality} is that the moderate values of data scale can generate significantly better nDCG results than that of smaller values, and the nDCG values have insignificant changes while data scale becomes much larger.

Based on this observation, we keep the underlaying environment unchanged and choose 1/16 as the scale factor value for our testbed in the following experiments, i.e. $ds=1/16$ and $nm=5$.

\subsubsection{Prediction model comparison} The second experiment is to evaluate the quality of different prediction models by comparing their rankings on the testbed. For each application, we set its time constraint first, and then construct the production system and the corresponding testbed to run the application independently under the constraint. After that, we collect samples on production system ($PS$) and on testbed ($TB$), and train prediction models, i.e. random forest (RF), gradient boosting decision tree (GBDT), and support vector regression (SVR), using the 10-fold cross-validation method on both $PS$ and $TB$. Table \ref{tab:testbed} lists the nDCG values on production system and testbed. We find that the nDCG values of all three models on testbed outperform those on production system. Specially, given the same time constraints, the nDCG values of RF on testbed obtain an average of 7.11\% improvement over them on production system, 5.28\% improvement for GBDT, and 5.68\% improvement for SVR. The results also indicate that the random-forest model achieves better performance than other two models.

\begin{table} \small
\centering \caption{The nDCG values of three different learning models}
\label{tab:testbed}
\begin{tabular}{I c |c I c|c|c|c I}\hline
Apps & \tabincell{c}{Time \\ Constraints} & \tabincell{c}{\# of \\ Samples} & \tabincell{c}{nDCG \\ (RF)} & \tabincell{c}{nDCG \\ (GBDT)} & \tabincell{c}{nDCG \\ (SVR)} \\ \hline

\multirow{2}*{\tabincell{c}{WC}} & \multirow{2}*{\tabincell{c}{1.65h}} & TB (216) & 0.9482 & 0.9026 & 0.8118 \\ \cline{3-6}

& & PS (31) & 0.9088 & 0.8909 & 0.8080 \\ \hline

\multirow{2}*{\tabincell{c}{BC}} & \multirow{2}*{\tabincell{c}{14.95h}} & TB (545) & 0.8412 & 0.7933 & 0.7501 \\ \cline{3-6}

&  & PS (118) & 0.7836 & 0.6942 & 0.6498 \\ \hline

\multirow{2}*{\tabincell{c}{KC}} & \multirow{2}*{\tabincell{c}{34.45h}} & TB (633) & 0.9038 & 0.8452 & 0.7851 \\ \cline{3-6}

&  & PS (94) & 0.7817 & 0.7116 & 0.7761 \\ \hline

\multirow{2}*{\tabincell{c}{LR}} & \multirow{2}*{\tabincell{c}{12.42h}} & TB (743) & 0.8714 & 0.8689 & 0.8077 \\ \cline{3-6}

&  & PS (92) & 0.7852 & 0.7947 & 0.7534 \\ \hline

\multirow{2}*{\tabincell{c}{SVM}} & \multirow{2}*{\tabincell{c}{64.4h}} & TB (267) & 0.9834 & 0.9355 & 0.8445 \\ \cline{3-6}

&  & PS (28) & 0.8404 & 0.9255 & 0.7264 \\ \hline

\multirow{2}*{\tabincell{c}{GBT}} & \multirow{2}*{\tabincell{c}{1.3h}} & TB (67) & 0.9052 & 0.8761 & 0.8783 \\ \cline{3-6}

&  & PS (23) & 0.8638 & 0.8702 & 0.7732 \\ \hline

\multirow{2}*{\tabincell{c}{PR}} & \multirow{2}*{\tabincell{c}{9.54h}} & TB (242) & 0.9032 & 0.8939 & 0.8491 \\ \cline{3-6}

&  & PS (59) & 0.8953 & 0.8585 & 0.8422 \\ \hline

\end{tabular}

\end{table}

\subsubsection{Hyperparameters} 
\begin{figure*}
\centering
\includegraphics[width=0.8\textwidth]{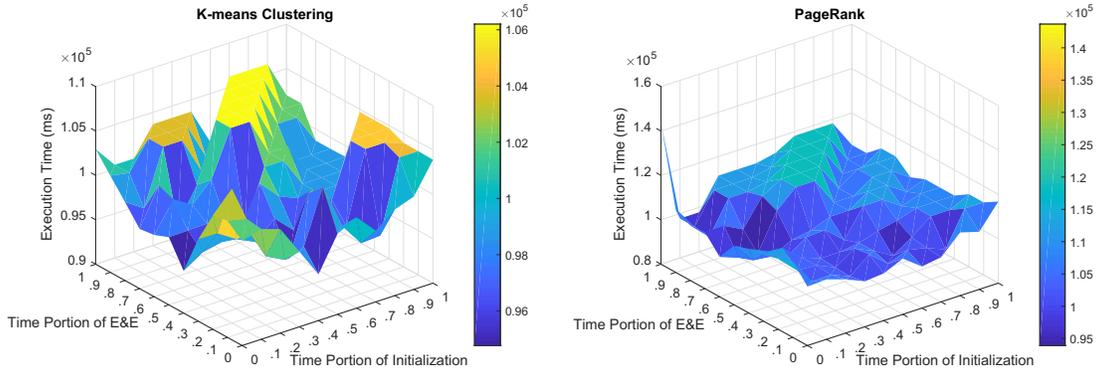}
\caption{Execution time results over different time ratios of three phases in AutoTune algorithm} \label{fig:paracomp}
\end{figure*}

The third experiment is to evaluate the performance variations over different time portions of initialization, exploration and exploitation (E\&E), and the best configuration selection phases in our AutoTune algorithm. For two applications \emph{K-Means Clustering (KC)} and \emph{PageRank (PR)}, we plot their execution time results over different time proportions of three phases in Figure \ref{fig:paracomp}, where $x$ axis denotes the initialization time portion from 0 to 1, y axis denotes the E\&E time portion from 0 to 1, and each point ($x, y, ET$) in the 3D plots represents the execution time under the best configuration generated by AutoTune algorithm. These performance results show that the time portions of three phases do affect the execution time results on both applications: about 10.77\% relative difference over the worst value on $KC$, and 34.62\% on $PR$. We can also observe the similar blue cross regions in the middle of these two 3D plots, which means that we can find the optimal configuration with a higher probability by balancing the time allocation to three phases.

\subsubsection{Execution time} Given a fixed time constraint for each application, we construct the testbed to run six different tuning algorithms plus default configuration independently. Table \ref{tab:differentresults} lists the execution time results (ms). As expected, the default configuration does not perform well. Our algorithm achieves an average of 63.70\% improvement over the default configurations. Furthermore, using testbed improves performance for all algorithms: random on testbed achieves an average of 9.57\% execution time improvement over only uses the production systems, BestConfig 15.18\% improvement, RFHOC 19.13\% improvement, Hyperopt 8.38\% improvement, SMAC 10.7\% improvement, and AutoTune 13.33\% improvement.

\begin{table*} \small
\centering \caption{Execution time results (ms) from different algorithms with fixed time constraints}
\label{tab:differentresults}
\begin{tabular}{I c|c|c I c|c|c|c|c|c|c I}\hline
Apps & \tabincell{c}{Time Constraints\\ (Running Times\footnotemark[1])} & \tabincell{c}{Target \\ Platform\footnotemark[2]} & Default\footnotemark[3] & Random & BestConfig & RFHOC & Hyperopt & SMAC & AutoTune \\ \hline

\multirow{2}*{\tabincell{c}{WC}} & \multirow{2}*{5.33h (100)} & TB & - & \tabincell{c}{169282 \\ (3.14\%)} & \tabincell{c}{171714 \\ (24.01\%)} & \tabincell{c}{172020 \\ (4.50\%)} & \tabincell{c}{170031 \\ (3.31\%)} & \tabincell{c}{168828 \\ (3.52\%)} & \tabincell{c}{161190 \\ (6.93\%)} \\ \cline{3-10}

& & PS & 217524 & 174780 & 225963 & 180126 & 175857 & 174991 & 173198 \\ \hline

\multirow{2}*{\tabincell{c}{BC}} & \multirow{2}*{12.6h (100)} & TB & - & \tabincell{c}{112406 \\ (3.34\%)} & \tabincell{c}{114643 \\ (5.39\%)}  & \tabincell{c}{114419 \\ (5.22\%)} & \tabincell{c}{109380 \\ (3.78\%)} & \tabincell{c}{108734 \\ (4.99\%)} & \tabincell{c}{105426 \\ (7.76\%)} \\ \cline{3-10}

& & PS & 425501 & 116293 & 121180 & 120722 & 113681 & 114439 & 114296 \\ \hline

\multirow{2}*{\tabincell{c}{KC}} & \multirow{2}*{18h (49)} & TB & - & \tabincell{c}{213174 \\ (16.58\%)} & \tabincell{c}{232508 \\ (17.76\%)} & \tabincell{c}{257460 \\ (20.84\%)} & \tabincell{c}{219063 \\ (17.87\%)} & \tabincell{c}{214776 \\ (19.15\%)} & \tabincell{c}{195545 \\ (16.05\%)} \\ \cline{3-10}

& & PS & 1301446 & 255550 & 282722 & 325247 & 266733 & 265648 & 232939 \\ \hline

\multirow{2}*{\tabincell{c}{LR}} & \multirow{2}*{6h (44)} & TB & - & \tabincell{c}{212231 \\ (14.80\%)} & \tabincell{c}{238246 \\ (13.66\%)} & \tabincell{c}{300004 \\ (33.33\%)} & \tabincell{c}{197013 \\ (11.26\%)} & \tabincell{c}{218886 \\ (20.47\%)} & \tabincell{c}{195380 \\ (12.27\%)} \\ \cline{3-10}

& & PS & 516677 & 249086 & 275951 & 449964 & 222016 & 275224 & 222694 \\ \hline

\multirow{2}*{\tabincell{c}{SVM}} & \multirow{2}*{24h (10)} & TB & - & \tabincell{c}{492229 \\ (13.32\%)} & \tabincell{c}{562842 \\ (34.71\%)} & \tabincell{c}{606586 \\ (61.97\%)} & \tabincell{c}{507454 \\ (7.34\%)} & \tabincell{c}{486047 \\ (10.78\%)} & \tabincell{c}{452801 \\ (15.55\%)} \\ \cline{3-10}

& & PS & 7880587 & 567873 & 862026 & 1595021 & 547657 & 544763 & 536207 \\ \hline

\multirow{2}*{\tabincell{c}{GBT}} & \multirow{2}*{6h (105)} & TB & - & \tabincell{c}{161352 \\ (5.70\%)} & \tabincell{c}{162568 \\ (0.67\%)} & \tabincell{c}{161137 \\ (3.56\%)} & \tabincell{c}{162546 \\ (3.77\%)} & \tabincell{c}{158569 \\ (4.40\%)} & \tabincell{c}{149186 \\ (10.68\%)} \\ \cline{3-10}

& & PS & 201714 & 171097 & 163664 & 167082 & 168916 & 165873 & 167025 \\ \hline

\multirow{2}*{\tabincell{c}{PR}} & \multirow{2}*{8.08h (50)} & TB & - & \tabincell{c}{81830 \\ (10.11\%)} & \tabincell{c}{89343 \\ (10.08\%)} & \tabincell{c}{97770 \\ (4.54\%)} & \tabincell{c}{77996 \\ (11.30\%)} & \tabincell{c}{80429 \\ (7.18\%)} & \tabincell{c}{74719 \\ (24.09\%)} \\ \cline{3-10}

& & PS & 328951 & 91032 & 99354 & 102416 & 87929 & 86647 & 98426 \\ \hline

\end{tabular}

\raggedright
\ \ \ \ \footnotemark[1]{Running Times: given the fixed time constraints, the approximate running times on production system with default configuration.}

\ \ \ \ \footnotemark[2]{Target Platform: where the algorithms work on and evaluate different configurations. TB = Testbed, PS = Production System.}

\ \ \ \ \footnotemark[3]{The performance of default configurations on testbed is not recorded for its unimportance.}

\end{table*}

\begin{figure}
\centering
\includegraphics[width=0.48\textwidth]{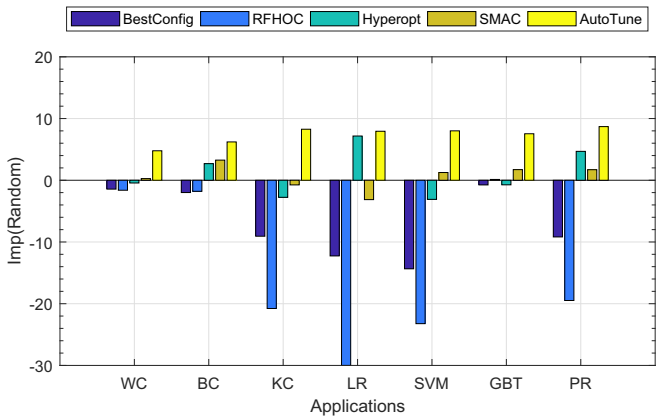}
\caption{Performance comparison among different algorithms using testbed} \label{fig:algocomp}
\end{figure}

Finally, we plot the overall execution time improvement percentage of BestConfig, RFHOC, Hyperopt, SMAC and AutoTune in Figure \ref{fig:algocomp}, using the random algorithm as the baseline. In the Figure \ref{fig:algocomp}, $x$-axis lists the seven different applications and $y$-axis represents the improvement percentage over the random algorithm. We observe that compared with the random algorithm, our approach achieves 4.8\%--8.7\% improvement among all applications. AutoTune achieves an average of 7.35\% improvement over Random, 14.35\% improvement over BestConfig, 22.79\% improvement over RFHOC, 6.28\% improvement over Hyperopt, and 6.73\% improvement over SMAC. We can conclude from Figure \ref{fig:algocomp} that AutoTune achieves stable and significant improvements compared with the other five algorithms. Another interesting observation from Figure \ref{fig:algocomp} is that the random search achieves surprisingly good results in our experiments. This is consistent with the findings of Bergstra and Bengio in \cite{bergstra2012random}.

\section{Conclusion and Future Work}\label{sec:conclusion}
In this paper, we propose AutoTune--an automatic configuration tuning system to optimize execution time for BDAFs. AutoTune constructs a smaller-scale testbed from the production system so that it can generate more samples, and thus train a better prediction model, under a given time constraint. Furthermore, the AutoTune algorithm selects a set of samples that can provide a wide coverage over the high-dimensional parameter space, and searches for more promising configurations using the trained prediction model.

It is of our future work to refine our testbed approach by supporting the automatic selection of the appropriate testbed settings given a scale factor and a resource constraint. We will also investigate the performance dynamics of BDAFs and design better approaches to account for such dynamics in our algorithm design. Last, we hope to integrate the proposed algorithm into the major Hadoop/Spark releases to support intelligent and automatic parameter tuning for big data analytics applications.


\bibliographystyle{IEEEtran}
\balance
\bibliography{autotune}

\end{document}